\newcommand{\Tr}{\mathop{\mathrm{Tr}}\nolimits}

\documentclass[pra,superscriptaddress,showpacs,twocolumn,eqsecnum]{revtex4}
\usepackage{amsfonts,amssymb,bm,graphicx}

\begin{document}

\title{Distance-based degrees of polarization
for a quantum field}

\author{A. B. Klimov}
\affiliation{Departamento de F\'{\i}sica,
Universidad de Guadalajara,
44420~Guadalajara, Jalisco, Mexico}

\author{L. L. S\'anchez-Soto}
\affiliation{Departamento de \'{O}ptica,
Facultad de F\'{\i}sica,
Universidad Complutense, 28040~Madrid, Spain}

\author{E. C. Yustas}
\affiliation{Departamento de \'{O}ptica,
Facultad de F\'{\i}sica,
Universidad Complutense, 28040~Madrid, Spain}

\author{J. S\"{o}derholm}
\affiliation{Institute of Quantum Science,
Nihon University, 1-8 Kanda-Surugadai,
Chiyoda-ku, Tokyo 101-8308, Japan}

\author{G. Bj\"{o}rk}
\affiliation{School of Communication and
Information Technology,
Royal Institute of Technology (KTH),
Electrum 229, SE-164 40 Kista, Sweden}

\date{\today}

\begin{abstract}
It is well established that unpolarized light
is invariant with respect to any SU(2) polarization
transformation. This requirement fully characterizes
the set of density matrices representing unpolarized
states. We introduce the degree of polarization of a
quantum state as its distance to the set of unpolarized
states. We use two different candidates of distance,
namely the Hilbert-Schmidt and the Bures metric,
showing that they induce fundamentally different
degrees of polarization. We apply these notions
to relevant field states and we demonstrate that
they avoid some of the problems arising with the
classical definition.
\end{abstract}


\maketitle

\section{Introduction}

The polarization properties of quantum
fields have received much attention over
the past few years, especially in the
single photon regime. The polarization
state is a robust characteristic, which
is relatively simple to manipulate without
inducing more than marginal losses. It is
thus hardly surprising that many outstanding
experiments in quantum optics, such as Bell
tests~\cite{Asp81,Kwi95}, quantum
tomography~\cite{Mar03} or quantum
cryptography~\cite{Ben92,Mul97}, have been
performed using polarization states.

The polarization  of a light beam in classical optics can be
elegantly visualized by resorting to the Poincar\'e sphere, and is
determined by the four Stokes
parameters~\cite{Sto52,Bor80,Sim90,Saa04}. These parameters
present unique advantages: they are easily measured, they can be
extended to the quantum domain, where the Stokes parameters become
the mean values of the Stokes operators, and, finally, they allow
us to classify the states of light according to a degree of
polarization~\cite{Jau76,Col70,Chi93,Alo99}.

This classical degree of polarization is simply the modulus of the
Stokes vector. While this affords a very intuitive image, it has
also serious drawbacks that can be traced back to the fact that
the Stokes parameters are proportional to the second-order
correlations of the field amplitudes. This may be sufficient for
most classical problems, but for quantum fields higher-order
correlations are crucial. For this reason, the Stokes parameters
do not distinguish between quantum states having remarkably
different polarization properties~\cite{Tse00,Tri00,Usa01}. For
example, the classical polarization degree of the state with
exactly one photon in each of the horizontally and vertically
polarization modes is zero, but it cannot be regarded as
unpolarized. These unwanted consequences call for alternative
measures.

Recently, Luis~\cite{Lui02} has brought in a challenging idea to
circumvent these difficulties. For the full characterization of
polarization, a probability distribution (obtained via the $Q$
function) is defined on the Poincar\'e sphere. To some extent, the
existence of such a probabilistic description is unavoidable in
quantum optics from the very beginning, since the Stokes operators
do not commute and thus no state can have a definite value of all
of them simultaneously (except the vacuum). In this framework, the
degree of polarization of a field state can be defined as the
distance from its associated distribution to the uniform
distribution corresponding to unpolarized light. We find this
proposal interesting, but semiclassical in nature. In addition,
we stress that the SU(2) $Q$ function does not connect manifolds
with different photon excitations, so the information it embodies
cannot be, in general, complete.

The question of what unpolarized light is has a relatively long
history~\cite{Pra71,Aga71,Leh96}. Today, there is a wide
consensus~\cite{SPIE,Wun03} in considering unpolarized light as
the light that is described by quantum states that are invariant
with respect to any SU(2) polarization transformation. It turns
out that this requirement fixes the density operators admissible
to represent unpolarized fields. It is then suggestive to look at
the degree of polarization as a distance from a given state to
this set of unpolarized states.

The notion of distance measure has been successfully used in
assessing nonclassicality~\cite{Hil87,Dod00,Mar02},
entanglement~\cite{Ved97}, quantum
information~\cite{Sch95,Bruk99,Sch99,Rud01,Chi00,Nie02,Ter04,Gil04,Mau05}
and localization~\cite{Maa88,And93,Mir98,Gnu01}, to cite only some
relevant examples. These measures are useful both when comparing
experiments with the corresponding theory and in comparing
different experiments. We hope that they will be soon agreed upon
by experimentalists and theorists alike.

In this paper, we connect the idea of distance with the problem of
assessing the polarization characteristics of a quantum field,
exploring a suitable definition that avoids  at least some of the
aforementioned difficulties that previous approaches based on
Stokes parameters encounter.

\section{SU(2) polarization structure and
invariance properties of quantum fields}

We assume a monochromatic plane wave propagating in the $z$
direction, whose electric field lies in the $xy$ plane. Under
these conditions, we are dealing with a two-mode field that can be
fully described by two complex amplitude operators. They are
denoted by $\hat{a}_H$ and $\hat{a}_V$, where the subscripts $H$
and $V$ indicate horizontally and vertically polarized modes,
respectively. The commutation relations of these operators are
standard:
\begin{equation}
\label{bccr}
[\hat{a}_j, \hat{a}_k^\dagger ] =
\delta_{jk} \, ,
\qquad j, k \in \{H, V \} \, .
\end{equation}
The Stokes operators are then defined as the
quantum counterparts of the classical variables,
namely~\cite{Jau76,Col70,Chi93,Alo99}
\begin{eqnarray}
\label{Stokop}
\hat{S}_0 = \hat{a}^\dagger_H \hat{a}_H +
\hat{a}^\dagger_V \hat{a}_V \, ,
\qquad
\hat{S}_1 = \hat{a}^\dagger_H \hat{a}_V +
\hat{a}^\dagger_V \hat{a}_H \, , \nonumber \\
& & \\
\hat{S}_2 = i ( \hat{a}_H \hat{a}^\dagger_V -
\hat{a}^\dagger_H \hat{a}_V ) \, ,
\qquad
\hat{S}_3 = \hat{a}^\dagger_H \hat{a}_H -
\hat{a}^\dagger_V \hat{a}_V \, , \nonumber
\end{eqnarray}
and their mean values are precisely the Stokes
parameters $(\langle \hat{S}_0 \rangle, \langle
\hat{\mathbf{S}} \rangle )$, where $\hat{\mathbf{S}}
= (\hat{S}_1, \hat{S}_2, \hat{S}_3)$. Using
the relation (\ref{bccr}), one immediately gets
that the Stokes operators satisfy the commutation
relations of angular momentum:
\begin{equation}
\label{crsu2}
[\hat{\mathbf{S}}, \hat{S}_0] = 0 \, ,
\qquad
[ \hat{S}_1, \hat{S}_2] = 2 i \hat{S}_3 \, ,
\end{equation}
and cyclic permutations. The noncommutability of
these operators precludes the simultaneous exact
measurement of their physical quantities. Among
other consequences, this implies that no field
state (leaving aside the two-mode vacuum) can
have definite nonfluctuating values of all the
Stokes operators simultaneously. This is
expressed by the uncertainty relation
\begin{equation}
(\Delta \hat{\mathbf{S}} )^2 =
(\Delta \hat{S}_1)^2 + (\Delta \hat{S}_2)^2
+ (\Delta \hat{S}_3)^2 \geq 2 \langle
\hat{S}_0 \rangle \, .
\end{equation}
Contrary to what happens in classical optics, the
electric vector of a monochromatic quantum field
never describes a definite ellipse~\cite{Lui02}.

In mathematical terms, a SU(2) (or linear) polarization
transformation is any transformation generated by the operators
$\hat{\mathbf{S}}$. It is well known~\cite{Yur86} that the
operator $\hat{S}_2$ is the infinitesimal generator of geometrical
rotations around the direction of propagation, whereas $\hat{S}_3$
is the infinitesimal generator of differential phase shifts
between the modes. As indicated by Eq.~(\ref{crsu2}), these two
operators suffice to generate all SU(2) polarization
transformations, which in experimental terms means that they can
be accomplished with a combination of phase plates and rotators
(that produce rotations of the electric field components around
the propagation axis)~\cite{SPIE}.

The standard definition of the degree of
polarization is~\cite{Bor80,Sim90,Saa04}
\begin{equation}
\mathcal{P}_{\mathrm{sc}} =
\frac{\sqrt{\langle
\hat{\mathbf{S}} \rangle^2}}
{\langle \hat{S}_0 \rangle} =
\frac{\sqrt{\langle \hat{S}_1 \rangle^2
+ \langle \hat{S}_2 \rangle^2 +
\langle \hat{S}_3 \rangle^2}}
{\langle \hat{S}_0 \rangle} \, ,
\end{equation}
where the subscript sc indicates that this is a semiclassical
definition, mimicking the form of the classical one. In the
semiclassical description it is implicitly assumed that
unpolarized light (i.~e., the origin of the Poincar\'e sphere) is
defined by the specific values~\cite{Kar93}
\begin{equation}
\langle \hat{S}_1 \rangle =
\langle \hat{S}_2 \rangle =
\langle \hat{S}_3 \rangle = 0 \, .
\end{equation}
Sometimes the extra requirement that the Stokes
parameters are temporally invariant is added to
make the definition even more stringent~\cite{Bar89}.
In any case, this conception has several flaws
that have been put forward before and give rise
to strange concepts such as that of quantum
states with ``hidden" polarization~\cite{Kly92}.
Actually, this notion leads to the paradoxical
conclusion that unpolarized light has a
polarization structure, which is latent when
the mean intensities are measured and detectable
when the noise intensities are measured~\cite{Kar95}.

If, as anticipated in the Introduction, one looks
at unpolarized light as field states that
remain invariant under any SU(2) polarization
transformation, then there is no more any ``hidden"
polarization. Any state satisfying this invariance
condition will also fulfill the classical definition
of an unpolarized state, but the converse is not true.
It has been shown~\cite{Pra71,Aga71,Leh96} that the
density operator of such ``quantum" unpolarized
states can be always written as
\begin{equation}
\label{denunpol}
\hat{\sigma} = \bigoplus_{N=0}^\infty
\lambda_N  \hat{\openone}_N \, ,
\end{equation}
where $N$ denotes the excitation manifold in which
there are exactly $N$ photons in the field. All
the coefficients $\lambda_N$ are real and nonnegative
and to ensure the unit-trace condition of the
density operator they must satisfy
\begin{equation}
\label{consl}
\sum_{N=0}^\infty (N+1) \lambda_N = 1 \, .
\end{equation}

In the following, the basis states of the excitation
manifold $N$ will be denoted as $|N, k \rangle =
| k \rangle_H \otimes |N - k \rangle_V$, $k = 0, 1,
\ldots, N$. These states span a SU(2) invariant
subspace of dimension $N+1$, and the generators
$\hat{\mathbf{S}}$ act therein according to
\begin{eqnarray}
\hat{S}_+ |N, k \rangle & = &
\sqrt{( k + 1 ) ( N - k)} |N, k +1 \rangle \, ,
\nonumber \\
\hat{S}_- | N, k \rangle & = &
\sqrt{k (N - k +1)} | N, k -1 \rangle \, , \\
\hat{S}_3 | N, k \rangle & = &
( k - N/2 ) | N, k \rangle \, , \nonumber
\end{eqnarray}
where $\hat{S}_\pm = \hat{S}_1 \pm i \hat{S}_2$.
These SU(2) invariant subspaces will play a
key role in the following.

\section{Quantum degree of polarization
as a distance}

As we have discussed before, measures of
nonclassicality have been defined as the
(minimum) distance to an appropriate set
representing classical states~\cite{Hil87,Dod00,Mar02}.
Similarly, the  minimum distance to the
(convex) set of separable states has
been used to introduce measures of
entanglement~\cite{Ved97}.  In the same
vein, we propose to quantify the degree
of polarization as
\begin{equation}
\label{DoP}
\mathbb{P} (\hat{\rho}) \propto
\inf_{\hat{\sigma} \in \mathcal{U}}
D(\hat{\rho} , \hat{\sigma} ) \, ,
\end{equation}
where $\mathcal{U}$ denotes the set of unpolarized states of the
form (\ref{denunpol}) and $D(\hat{\rho}, \hat{\sigma} )$ is any
measure of distance (not necessarily a metric) between the two
density matrices $\hat{\rho}$ and $\hat{\sigma}$, such that
$\mathbb{P} (\hat{\rho})$ satisfies some requirements motivated by
both physical and mathematical concerns. The constant of
proportionality in Eq.~(\ref{DoP}) must be chosen in such a way
that $\mathbb{P}$ is normalized to unity, i.e., $
\inf_{\hat{\rho}} \mathbb{P} (\hat{\rho}) = 1$.

In Ref.~\cite{Gil04} a check list of six simple,
physically-motivated criteria that should be
satisfied by any good measure of distance between
quantum processes can be found. For our problem, we
impose the following two conditions:
\begin{description}
    \item[(C1)] $\mathbb{P} (\hat{\rho}) = 0$
    iff $\hat{\rho}$ is unpolarized.

    \item[(C2)] Energy-preserving unitary
    transformations $\hat{U}_E$ leave $\mathbb{P}
    (\hat{\rho})$ invariant; that is,
    $\mathbb{P} (\hat{\rho}) = \mathbb{P}
    (\hat{U}_E \hat{\rho} \hat{U}_E^\dagger).$
\end{description}
The first condition is to some extent trivial: it
ensures that unpolarized and only unpolarized
states have a zero degree of polarization. The
second takes into account that the requirement
that an unpolarized state is invariant under any
SU(2) polarization transformation makes it also
invariant under any energy-preserving unitary
transformation~\cite{Seh05}: these include not
only the transformations generated by
$\hat{\mathbf{S}}$, but also those generated
by $\hat{S}_0$, which, in technical terms,
corresponds to the group U(2)~\cite{Wun03}.

It is clear that there are numerous nontrivial
choices for $D(\hat{\rho} , \hat{\sigma})$ (by
nontrivial we mean that the choice is not a
simple scale transformation of any other
distance). None of them could be said to be
more important than any other \textit{a priori},
but the significance of each candidate would
have to be seen through physical assumptions.
To illustrate this point further, let us take
an extreme example fulfilling the previous
conditions~\cite{Ved97}. Define the discrete
distance
\begin{equation}
D_{\mathrm{dis}} (\hat{\rho} , \hat{\sigma}) =
\left \{
\begin{array}{ll}
1 \, , & \qquad
\hat{\rho} \neq \hat{\sigma} \, , \\
& \\
0 \, , & \qquad
\hat{\rho} = \hat{\sigma} \, .
\end{array}
\right .
\end{equation}
If the degree of polarization is computed
using this distance, we have
\begin{equation}
\mathbb{P}_{\mathrm{dis}} (\hat{\rho} ) =
\left \{
\begin{array}{ll}
1\, , & \qquad
\hat{\rho} \notin \mathcal{U} \, , \\
& \\
0 \, , & \qquad
\hat{\rho} \in \mathcal{U} \, .
\end{array}
\right .
\end{equation}
This therefore tells us only if a given
state $\hat{\rho}$ is unpolarized or not.

There are authors demanding that $D (\hat{\rho} , \hat{\sigma})$
is a metric~\cite{Gil04}. This requires three additional
properties: (1)~Positiveness: $D (\hat{\rho}, \hat{\sigma}) \ge 0$
and $D (\hat{\rho}, \hat{\sigma}) = 0$ iff $\hat{\rho} =
\hat{\sigma}$. (2)~Symmetry: $D (\hat{\rho} , \hat{\sigma}) = D
(\hat{\sigma}, \hat{\rho})$. (3)~Triangle inequality: $D
(\hat{\rho} , \hat{\tau}) \leq D (\hat{\rho} , \hat{\sigma}) + D
(\hat{\sigma}, \hat{\tau})$. These are quite reasonable
properties, since most distances used in quantum mechanics are
based on an inner product and so they automatically fulfill them.
However, there exist pertinent examples in which $D$ is not a
metric. For example, the quantum relative
entropy~\cite{Weh78,Ohy89,Hia91,Don86}
\begin{equation}
D_S (\hat{\rho}, \hat{\sigma}) =
S (\hat{\rho} || \hat{\sigma}) =
\Tr [ \hat{\rho} (\ln \hat{\rho} -
\ln \hat{\sigma} ) ]
\end{equation}
is not symmetric and does not satisfy the triangle inequality.
Nevertheless, it generates a valuable measure of entanglement, and
the corresponding degree of polarization satisfies both C1 and C2.

However, for a detailed analysis we will consider the
Hilbert-Schmidt metric
\begin{equation}
D_{\mathrm{HS}} (\hat{\rho}, \hat{\sigma}) = || \hat{\rho} -
\hat{\sigma}||_{\mathrm{HS}}^2 = \Tr [( \hat{\rho} -
\hat{\sigma})^2 ] \, ,
\end{equation}
which has been previously studied in the contexts of
entanglement~\cite{Wit99,Oza00,Ber02}. Since $D_{\mathrm{HS}}
(\hat{\rho}, \hat{\sigma})$ is a metric, condition C1 is
satisfied. It follows from the unitary invariance of the
Hilbert-Schmidt metric that also C2 is satisfied.

According to our general strategy stated in the
definition~(\ref{DoP}), for a given state $\hat{\rho}$  we should
find the unpolarized state $\hat{\sigma}$ that minimizes the
distance
\begin{equation}
\label{HSdist}
D_{\mathrm{HS}} ( \hat{\rho} , \hat{\sigma})
=  \Tr ( \hat{\rho}^2 ) + \Tr (\hat{\sigma}^2 )
- 2 \Tr (\hat{\rho} \hat{\sigma} ) \, .
\end{equation}
If we take into account that the purity of an
unpolarized state is
\begin{equation}
\label{UnpolPur}
\Tr ( \hat{\sigma}^2 ) =
\sum_{N=0}^\infty (N + 1) \lambda_N^2 \, ,
\end{equation}
we easily get
\begin{equation}
D_{\mathrm{HS}} ( \hat{\rho} , \hat{\sigma})
= \Tr (\hat{\rho}^2 )  +
\sum_{N=0}^\infty [ (N + 1) \lambda_N^2 -
2 p_N \lambda_N ] \, ,
\end{equation}
where $p_N$ is the probability distribution of
the total number of photons
\begin{equation}
p_N = \sum_{k=0}^N \rho_{Nk,Nk} \, ,
\end{equation}
and $\rho_{Nk, N^\prime k^\prime} = \langle N, k | \hat{\rho} |
N^\prime, k^\prime \rangle$ denotes the matrix elements of the
density operator $\hat{\rho}$ between SU(2) invariant subspaces.
Now, it is easy to obtain the coefficients $\lambda_N$ that
minimize this distance. The calculation is direct and the result
is
\begin{equation}
\label{lNopt}
\lambda_N = \frac{p_N}{N + 1} \, .
\end{equation}
The density operator $\hat{\sigma}_{\mathrm{opt}}
\in \mathcal{U}$ with these optimum coefficients $
\lambda_N$ satisfies the constraint (\ref{consl}),
and hence minimizes the distance (\ref{HSdist}).
Note that $\hat{\sigma}_{\mathrm{opt}}$ can be
written as
\begin{equation}
\label{dep}
\hat{\sigma}_{\mathrm{opt}} =
\sum_{N=0}^\infty p_N \
\hat{\sigma}_{\mathrm{opt}}^{(N)} \, ,
\end{equation}
with
\begin{equation}
\hat{\sigma}_{\mathrm{opt}}^{(N)}
= \frac{1}{N+1}
\sum_{k = 0}^{N} |N, k \rangle
\langle N, k |  \, .
\end{equation}

With all this in mind, we can define the
Hilbert-Schmidt degree of polarization by
\begin{equation}
\label{PHS}
\mathbb{P}_{\mathrm{HS}} (\hat{\rho}) =
\Tr ( \hat{\rho}^2 ) -
\sum_{N=0}^\infty \frac{p_N^2}{N + 1} \, ,
\end{equation}
which is determined not only by the purity
$0  < \Tr ( \hat{\rho}^2 ) \leq 1$ (as it
happens for other measures~\cite{Hel87}),
but also by the distribution of the number
of photons $p_N$.  Although the maximum
Hilbert-Schmidt distance between two density
operators is $2$, the minimum distance to
an unpolarized state is normalized to unity.

Using Eqs.~(\ref{UnpolPur}) and (\ref{lNopt}),
the Hilbert-Schmidt degree of polarization can
be recast as
\begin{equation}
\mathbb{P}_{\mathrm{HS}} (\hat{\rho}) =
\text{Tr} (\hat{\rho}^2 ) -
\Tr  ( \hat{\sigma}_{\mathrm{opt}}^2) \, ,
\end{equation}
which makes it easy to verify that it
vanishes only for unpolarized states,
in agreement with the condition C1.

It has been shown~\cite{Wit99,Oza00,Ber02} that the
Hilbert-Schmidt distance is not monotonically decreasing under
every completely positive trace-preserving map (what is called the
CP nonexpansive property). This has motivated that the quantum
information community has identified the fidelity as a
particularly important alternative approach to the definition of a
distance measure for states~\cite{Nie00}.

In consequence, as our second candidate of distance we will employ
the fidelity (or Uhlmann transition probability) \cite{Uhl76}
\begin{equation}
\label{fid}
F (\hat{\rho}, \hat{\sigma}) =
[ \Tr ( \hat{\sigma}^{1/2} \, \hat{\rho} \,
\hat{\sigma}^{1/2})^{1/2} ]^2 \, .
\end{equation}
A word of caution is necessary here. There is
an ambiguity in the literature: both the
quantity~(\ref{fid}) and its square root have
been referred to as the fidelity. The reader
should take this into account when comparing
different sources.

The fidelity has many attractive properties. First, it is
symmetric in its arguments $F (\hat{\rho}, \hat{\sigma}) = F
(\hat{\sigma}, \hat{\rho})$, a fact that is not obvious from
Eq.~(\ref{fid}), but which follows from other equivalent
expressions. It can also be shown that $0 \le F (\hat{\rho},
\hat{\sigma}) \le 1$, with equality in the second inequality iff
$\hat{\rho} = \hat{\sigma}$. This means that the fidelity is not a
metric as such, but serves rather as a generalized measure of the
overlap between two quantum states. A common way of turning it
into a metric is through the Bures metric
\begin{equation}
D_{\mathrm{B}}( \hat{\rho}, \hat{\sigma}) =
2 [1 - \sqrt{F (\hat{\rho},\hat{\sigma})}] \, .
\end{equation}
The origin of this distance can be seen
intuitively by considering the case when
$\hat{\rho}$ and $\hat{\sigma}$ are both
pure states. The Bures metric is just the
Euclidean distance between the two pure
states, with respect to the usual norm
on the state space.

Since the larger the fidelity $F (\hat{\rho},
\hat{\sigma})$, the smaller the Bures distance
$D_{\mathrm{B}}( \hat{\rho}, \hat{\sigma})$,
we can define the Bures degree of polarization
as
\begin{equation}
\label{defPB}
\mathbb{P}_{\mathrm{B}} (\hat{\rho}) =
1 - \sup_{\hat{\sigma} \in \mathcal{U}}
\sqrt{F (\hat{\rho},\hat{\sigma})} \, .
\end{equation}
An alternative definition would be $1 - \sup_{\hat{\sigma} \in
\mathcal{U}} F (\hat{\rho},\hat{\sigma})$, which arises naturally
in the context of quantum computation~\cite{Gil04}. It is clear
that these definitions order the states $\hat{\rho}$ in the same
way. Unfortunately, we cannot find a general expression of the
unpolarized state $\hat{\sigma}$ that gives the maximum fidelity.
Such a task must be performed case by case and will be illustrated
with some selected examples in the next Section. Let us conclude
this Section by noting that the fidelity can be used to define a
measure of entanglement \cite{Ved97}, whereas this is not the case
with the Hilbert-Schmidt metric \cite{Oza00}. Since the
unpolarized states are separable, $\mathbb{P}_{\mathrm{B}}
(\hat{\rho})$ thus gives an upper bound on the entanglement of
$\hat{\rho}$.

\section{Examples}

It is clear from Eq.~(\ref{PHS}) that all pure
$N$-photon states have the same Hilbert-Schmidt
degree of polarization. For such states, denoted
$|\Psi^{(N)} \rangle$, we have
\begin{equation}
\label{PHSNs}
\mathbb{P}_{\mathrm{HS}}( |\Psi^{(N)}
\rangle ) = \frac{N}{N + 1} \, .
\end{equation}
The Bures degree of polarization for these
states can also be readily found:
\begin{equation}
\label{PBps}
\mathbb{P}_{\mathrm{B}} ( | \Psi^{(N)}
\rangle ) = 1 - \frac{1}{\sqrt{N + 1}} \, .
\end{equation}
The vacuum is the only unpolarized state, in agreement with
condition C1. Note also that the expressions (\ref{PHSNs}) and
(\ref{PBps}) apply, e.~g., to the states $| n \rangle_H \otimes |n
\rangle_V$. Since for them $\langle \hat{\mathbf{S}} \rangle = 0$,
classically they would be unpolarized for every $n$ (that is,
$\mathcal{P}_{\mathrm{sc}} = 0$, even in the limit $n \gg 1$). In
our distance-based approach, the degree of polarization is a
function of all moments of the Stokes operators and not only of
the first one, as it happens for $\mathcal{P}_{\mathrm{sc}}$,
which causes this quite different behavior. We also observe that
all these states lying in the $N+1$-dimensional invariant subspace
satisfy $\mathbb{P} \rightarrow 1$ when their intensity is
increased.

Next, we define the diagonal states as those
that can be expressed as
\begin{equation}
\label{diag} \hat{\rho}_{\mathrm{diag}} = \sum_{N=0}^\infty
\sum_{k=0}^N p_{N k} | \Psi_k^{(N)} \rangle \langle \Psi_k^{(N)} |
\, ,
\end{equation}
where we let $p_{N k} \geq p_{N k+1}$, for all $k < N$, and $\{ |
\Psi_k^{(N)} \rangle \}_{k=0}^N$ is an arbitrary orthonormal basis
in excitation manifold $N$. It then follows from C2 that any two
diagonal states whose probability distribution $\{ p_{N k}
\}_{k=0}^N$ coincide, must have the same degree of polarization.
For any diagonal state, we have
\begin{eqnarray}
\mathbb{P}_{\mathrm{HS}}
(\hat{\rho}_{\mathrm{diag}}) & = &
\sum_{N=0}^\infty \sum_{k=0}^N p_{N k}^2 -
\sum_{N=0}^\infty \frac{p_N^2}{N + 1}
\nonumber \\
& \leq & \sum_{N=0}^\infty
\frac{N p_N^2}{N + 1} \, .
\end{eqnarray}

To deal with the Bures degree of polarization for
this example, we first note that
\begin{equation}
\sqrt{F (\hat{\rho}_{\mathrm{diag}},
\hat{\sigma})} = \sum_{N=0}^\infty \sum_{k=0}^N
\sqrt{\lambda_N \, p_{N k}} =
\sum_{N=0}^\infty s_N \, \sqrt{\lambda_N} \, ,
\end{equation}
where
\begin{equation}
s_N = \sum_{k=0}^N \sqrt{p_{N k}} \, .
\end{equation}
The extremal points of the fidelity are
then determined by
\begin{equation}
\frac{s_N}{2 \sqrt{\lambda_N}} - \mu (N + 1) = 0 \, ,
\end{equation}
where $\mu$ is a Lagrange multiplier that takes
into account the constraint (\ref{consl}). Solving
for $\lambda_N$ and imposing again (\ref{consl})
to fix the value of $\mu$, the optimum parameters
$\lambda_N$ are found to be
\begin{equation}
\lambda_N =
\frac{s_N^2}{\displaystyle
(N + 1)^2 \sum_{k=0}^\infty
\frac{s_k^2}{k + 1}} \, .
\end{equation}
In this way, we finally arrive  at
\begin{equation}
\label{PBdiag}
\mathbb{P}_{\mathrm{B}}
(\hat{\rho}_{\mathrm{diag}} )
=  1 - \sqrt{\sum_{N=0}^\infty
\frac{s_N^2}{N + 1}} \, .
\end{equation}
One can easily prove that
\begin{equation}
\sqrt{p_N} \leq s_N \leq
\sqrt{(N + 1) p_N} \, ,
\end{equation}
so we have the bound
\begin{equation}
\mathbb{P}_{\mathrm{B}}
(\hat{\rho}_{\mathrm{diag}}) \leq  1 -
\sqrt{\sum_{N=0}^\infty \frac{p_N}{N + 1}}
< 1 \, .
\end{equation}
It is clear that $\sum_{N=0}^\infty p_N (N + 1)^{-1} \rightarrow
0$ is necessary for both $\mathbb{P}_{\mathrm{B}}
(\hat{\rho}_{\mathrm{diag}})$ and $\mathbb{P}_{\mathrm{HS}}
(\hat{\rho}_{\mathrm{diag}})$ to approach unity. The latter also
requires the purity to approach unity,  whereas this is not
necessary in order to have $\mathbb{P}_{\mathrm{B}}
(\hat{\rho}_{\mathrm{diag}}) \rightarrow 1$.

As another relevant example, let us consider the
case in which both modes are in (quadrature) coherent
states. The product of two quadrature coherent
states, which we shall denote by $| \alpha_H,
\alpha_V \rangle$, can be expressed as a Poissonian
superposition of SU(2) coherent states~\cite{Atk71}
\begin{equation}
| \alpha_H, \alpha_V \rangle =
\sum_{N=0}^\infty
p_N  |N, \theta, \phi \rangle \, ,
\end{equation}
where $p_N$ is the Poissonian factor
\begin{equation}
p_N = e^{- \bar{N}/2} \frac{ \bar{N}^{N/2}}{\sqrt{N!}} \,
\end{equation}
and $\bar{N} = |\alpha_H|^2 + |\alpha_V|^2$
is the average number of excitations. The SU(2)
coherent states are defined as~\cite{Per86}
\begin{eqnarray}
|N, \theta, \phi \rangle & = &
\sum_{k=0}^N
\left (
\begin{array}{c}
N \\
k
\end{array}
\right )^{1/2}
\left ( \sin \frac{\theta}{2} \right)^{N-k}
\left ( \cos \frac{\theta}{2} \right)^k
\nonumber \\
& \times & e^{-i k \phi} \  | N, k \rangle \, ,
\end{eqnarray}
and the state parameters are connected by the
relations
\begin{equation}
\alpha_H = e^{-i\phi /2} \sqrt{\bar{N}}
\sin \frac{\theta }{2} \, ,
\qquad
\alpha_V = e^{i\phi /2} \sqrt{\bar{N}}
\cos \frac{\theta }{2} \, .
\end{equation}

Taking into account that
\begin{equation}
\sum_{N=0}^\infty \frac{p_N^2}{N+1}
= \frac{\mathrm{I}_1 ( 2 \bar{N} )}{\bar{N}}
e^{-2 \bar{N}} \, ,
\end{equation}
where $\mathrm{I}_1 (z)$ is the modified Bessel
function, Eq.~(\ref{PHS}) reduces to
\begin{equation}
\mathbb{P}_{\mathrm{HS}} = 1 -
\frac{\mathrm{I}_1 (2 \bar{N})}
{\bar{N}} e^{-2 \bar{N}} \, .
\end{equation}
When $N \gg 1$ we can retain the first term
in the asymptotic expansion of $\mathrm{I}_1 (z)$
to obtain
\begin{equation}
\mathbb{P_{\mathrm{HS}}} \simeq
 1- \frac{1}{2\sqrt{\pi }
\bar{N}^{3/2}} \, .
\end{equation}
This tends again to unity. However, one may have expected a
$\bar{N}^{-1}$ behavior for coherent states, while the scaling is
$\bar{N}^{-3/2}$.

One can ask if the Hilbert-Schmidt and Bures measures order some
pairs of states differently. In the Appendix we show that this is
indeed the case, and the induced degrees of polarization are
therefore fundamentally different.

\section{Conclusion}

In conclusion, we have shown that quantum optics entails
polarization states that cannot be suitably described by the
(semi)classical formalism based on Stokes parameters. We have
advocated the use of a degree of polarization based on an
appropriate distance to the set of unpolarized states. Such a
definition is closely related to other recent proposals in
different areas of quantum optics and is well behaved even in
cases where the classical formalism fails.

\begin{acknowledgments}
We would like to acknowledge useful discussions
with G. M. D'Ariano and A. W\"{u}nsche. G. B.
acknowledges financial support from STINT and
from the Swedish Research Council.
\end{acknowledgments}

\bigskip

\appendix*

\section{Proof of the definitions' different ordering of states}


In this appendix, we will show that the Hilbert-Schmidt and the
Bures distances induce fundamentally different degrees of
polarization. To this end, we consider the states
\begin{equation}
\label{rhoN1N2} \hat{\rho}_{N_1 N_2} = \sum_{j, k=1}^2 \rho_{j k}
| \Psi^{(N_j)} \rangle \langle \Psi^{(N_k)} | \, ,
\end{equation}
where $| \Psi^{(N_1)} \rangle$ and $| \Psi^{(N_2)} \rangle$ are
orthogonal pure states with $N_1$ and $N_2$ photons, respectively.
We here assume that $N_1 \neq N_2$, and note that $\hat{\rho}_{N
N}$ is a diagonal state of the form (\ref{diag}). To simplify
calculations, we shall use the notation
\begin{equation}
p = \rho_{11} \, ,
\qquad
1 - p = \rho_{22} \, ,
\qquad
q = \rho_{12} = \rho_{21}^\ast \, .
\end{equation}
The states $| \Psi^{(N_1)} \rangle$ and $| \Psi^{(N_2)}$ then
correspond to $p=0$ and $p=1$, respectively, and the purity
becomes
\begin{equation}
\Tr ( \hat{\rho}_{N_1 N_2}^2 ) =
p^2 + (1 - p)^2 + 2 |q|^2 \, .
\end{equation}
We note in passing that
$1 - 2 p (1-p) \leq
\Tr ( \hat{\rho}_{N_1 N_2}^2 ) \leq 1$ and
$0 \leq |q|^2 \leq p (1-p)$.

In the basis $( | \Psi^{(N_1)} \rangle,
| \Psi^{(N_2)} \rangle)$, we can write
\begin{equation}
\hat{\sigma}^{1/2} \hat{\rho}
\hat{\sigma}^{1/2} =
\left (
\begin{array}{cc}
\lambda_{N_1} p &
\sqrt{\lambda_{N_1} \lambda_{N_2}} q \\
&  \\
\sqrt{\lambda_{N_1} \lambda_{N_2}} q^\ast &
\lambda_{N_2} (1-p)
\end{array}
\right ) \, .
\end{equation}
Since the eigenvalues of this matrix are
\begin{eqnarray}
\label{chipm}
\chi_\pm & = & \frac{1}{2}
\left \{ \lambda_{N_1} p +
\lambda_{N_2} (1 - p) \right .
\nonumber \\
& \pm & \left . \sqrt{[\lambda_{N_1} p - \lambda_{N_2} (1 - p)]^2
+ 4 \lambda_{N_1} \lambda_{N_2} |q|^2} \right \}  \, ,
\end{eqnarray}
the fidelity can be expressed as
\begin{eqnarray}
\label{F2}
F (\hat{\rho}, \hat{\sigma}) & = &
\chi_+ + 2 \sqrt{\chi_+ \chi_-} + \chi_- =
\lambda_{N_1} p + \lambda_{N_2} (1 - p)
\nonumber \\
& & \nonumber \\
& + &  2 \sqrt{\lambda_{N_1} \lambda_{N_2}
[p (1 - p) - |q|^2]} \, .
\end{eqnarray}
For any fixed $\lambda_{N_1}$, $\lambda_{N_2}$, and $p$, the
fidelity decreases as $|q|^2$ increases. This could be expected,
since the unpolarized states do not have any off-diagonal
elements.

The restriction (\ref{consl}) implies for
this problem that
\begin{equation}
\label{lambdaRelNondegenerate}
\lambda_{N_2} =
\frac{1 - (N_1 + 1) \lambda_{N_1}}{N_2 + 1} \, .
\end{equation}
In consequence, the coefficients that optimize
the fidelity are determined by
\begin{eqnarray}
\label{supfid}
\lefteqn{\frac{\partial F}
{\partial \lambda_{N_1}} = 0 = p -
\frac{(1+N_1) (1-p)}{1+N_2}}
&  & \nonumber \\
& + &  [1 - 2 \lambda_{N_1} (1+N_1)]
\sqrt{\frac{p (1-p) - |q|^2}{\lambda_{N_1}
[1 - \lambda_{N_1} (1+N_1)] (1+N_2)}} \, .
\nonumber \\
\end{eqnarray}

\begin{figure}
\centering
\resizebox{0.85\columnwidth}{!}{\includegraphics{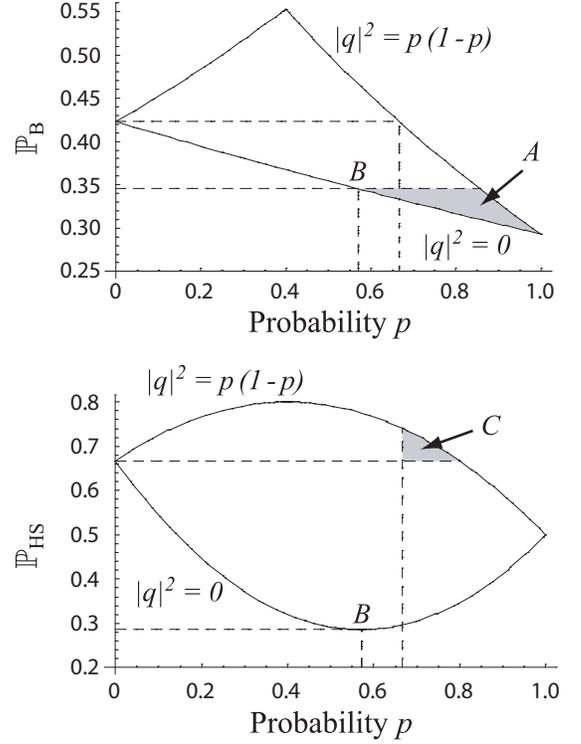}}
\caption{Polarization degrees for two-dimensional
states with $N_1 = 1$ and $N_2 = 2$. For any given $p$,
the maximum and minimum fidelities are given by $|q|^2 =
p (1-p)$ and $|q|^2 = 0$, respectively. Region $A$
corresponds to states satisfying $p > 4/7$ and
$\mathbb{P}_{\mathrm{B}} < 1 - \sqrt{3/7}$. For any state
$\hat{\rho}_A$ in this region,we have $\mathbb{P}_{\mathrm{B}}
(\hat{\rho}_A) < \mathbb{P}_{\mathrm{B}} (\hat{\rho}_B)$
while $\mathbb{P}_{\mathrm{HS}} (\hat{\rho}_A) >
\mathbb{P}_{\mathrm{HS}} (\hat{\rho}_B)$, where the
state $\hat{\rho}_B$ is characterized by $p = 4/7$ and
$|q|^2 = 0$. In region $C$, the states satisfy $p > 2/3$ and
$\mathbb{P}_{\mathrm{HS}} > 2/3$. For any such state
$\hat{\rho}_C$, we have $\mathbb{P}_{\mathrm{HS}} (\hat{\rho}_C) >
\mathbb{P}_{\mathrm{HS}} (|\Psi^{(2)} \rangle )$ and
$\mathbb{P}_{\mathrm{HS}} (\hat{\rho}_C) <
\mathbb{P}_{\mathrm{HS}} (| \Psi^{(2)} \rangle )$, where
$|\Psi^{(2)} \rangle$ is the (arbitrary) pure two-photon
state corresponding to $p = 0$.}
\end{figure}

We first consider pure states, for which $|q|^2 = p (1 - p)$.
Choosing $\lambda_{N_1}$ according to
\begin{equation}
\displaystyle
\begin{array}{ll}
\label{ArbitrarySol}
\lambda_{N_1} = 0 \, , &
\qquad \displaystyle
p < \frac{1+N_1}{2+N_1+N_2} \, , \\
& \\
\displaystyle
0 \leq \lambda_{N_1} \leq \frac{1}{1+N_1} \, , &
\qquad \displaystyle
p = \frac{1+N_1}{2+N_1+N_2} ,  \\
&  \\
\displaystyle
\lambda_{N_1} = \frac{1}{1 + N_1} \, , &
\qquad \displaystyle
p > \frac{1+N_1}{2+N_1+N_2} \, ,
\end{array}
\end{equation}
then maximizes the fidelity:
\begin{equation}
\sup_{\hat{\sigma} \in \mathcal{U}}
F (\hat{\rho}, \hat{\sigma})  =
\left \{
\begin{array}{ll}
\displaystyle
\frac{1-p}{1+N_2} \, , &
\qquad \displaystyle
p \leq \frac{1+N_1}{2+N_1+N_2} \, , \\
& \\
\displaystyle
\frac{p}{1+N_1} \, , &
\qquad \displaystyle
p \geq \frac{1+N_1}{2+N_1+N_2} \, .
\end{array}
\right .
\end{equation}

On the other hand, when $|q|^2 \neq p (1 - p)$ ($0 < p < 1$), the
solution of Eq.~(\ref{supfid}) is
\begin{eqnarray}
\label{NondegLambdaSol}
\lefteqn{\lambda_{N_1} =
\frac{1}{2 (N_1 + 1)}
\Biggl ( 1 - }   & &  \nonumber \\
& & \times \left .
\frac{(1 + N_1) (1 - p) - (1 + N_2) p}
{\sqrt{[1 + N_1 (1-p) + N_2 p]^2 -
4 (1 + N_1) (1 + N_2) |q|^2}}
\right )  \, . \nonumber  \\
\end{eqnarray}
Depending on the parameters, this solution can take any value in
the interval $0 < \lambda_{N_1} < 1/(1+N_1)$. In fact, one can
check that the choice (\ref{NondegLambdaSol}) gives the closest
unpolarized state. Combining Eqs.~(\ref{F2}),
(\ref{lambdaRelNondegenerate}), and (\ref{NondegLambdaSol}), thus
allows one to obtain the fidelity and hence
$\mathbb{P}_{\mathrm{B}}$.

In Fig.~1, we have plotted the Hilbert-Schmidt and Bures
degree of polarization for some two-dimensional states.
From the explanation in the caption, we see that the
two measures order some pairs of states differently.
The Hilbert-Schmidt and Bures distances thus induce
two fundamentally different degrees of polarization.

\end{document}